\documentstyle[11pt,mrs2001,epsfig]{article}

\def \hmpc{\rm{h^{-1}Mpc}}
\begin{document}

\title{THE LUMINOSITY DEPENDENCE OF CLUSTERING}

\author{E. HAWKINS}
\affil{School of Physics and Astronomy, University of Nottingham, Nottingham NG7 2RD, UK.}

\begin{abstract}
The PSCz redshift survey and the ongoing Two-Degree Field Galaxy
Redshift Survey (2dFGRS) are used to look for luminosity dependence of
galaxy clustering by measuring the two point correlation functions for
sub-samples of the surveys.

An analysis of the PSCz survey finds that galaxies with cooler colours
are more strongly clustered than warmer galaxies. Between $1$ and
$10~\hmpc$ it is found that the redshift-space correlation function,
$\xi(s)$, is a factor of $\sim 1.5$ larger for the cooler
galaxies. This is consistent with the suggestion that hotter galaxies
have higher star formation rates and correspond to later type galaxies
which are known to be less clustered than earlier types. There is a
very weak luminosity dependence of clustering with the more luminous
galaxies being less clustered than fainter galaxies, but the trend has
a low statistical significance.

Results from the 2dFGRS reveal a strong dependence of clustering on
intrinsic galaxy luminosity. The most luminous galaxies are more
strongly clustered by a factor $\sim 2.5$ than L$^*$ galaxies, in
qualitative agreement with analytic models of galaxy formation.
\end{abstract}

\section{Introduction}
Ever since the first galaxy surveys it has been clear that the galaxy
distribution is not uniform and that the measured clustering is
dependent on the properties of the sample in question. Differential
clustering as a function of morphological type, colour, luminosity and
wavelength probed have all been measured and imply the existence of
biases between the distribution of galaxies and mass. It is this
complex distribution of galaxies which galaxy formation models have
tried to reproduce. In the era of new large redshift surveys it is now
possible to test to much higher accuracy the large scale structure
predictions of these models. This talk concentrates on the luminosity
dependence of clustering which is perhaps the most controversial of
these measurements.

Typical galaxy formation models start with a smoothed random density
field, which is left to evolve. It is assumed that galaxies will form
at the peaks of this density field and that the mass of the galaxy is
related to the peak height.  If we then assume that galaxy luminosity
is related to galaxy mass we come to the conclusion that brighter
galaxies are found at the higher peaks. As higher peaks are more
strongly clustered \cite{ref1}, more luminous galaxies should also be
more strongly clustered. Hence these galaxy formation models predict
the amount of variation in galaxy clustering as a function of absolute
magnitude that we should observe. They all show a strong variation
over a wide range of magnitudes with a large increase of clustering
strength for the very bright galaxies \cite{benoist}. Although the
models do disagree on the exact amount of variation, this effect
should be large enough to be observable and the observations may allow
us to select the more realistic models.

Many previous studies have claimed to have observed this effect
(e.g. \cite{benoist}) but there have also been plenty of
non-detections (e.g. \cite{lo}). This controversy is likely due to the
limited sizes and small number of galaxies in past surveys which have
restricted the accuracy of the measurements. I use two recent, large,
but very different redshift surveys to try and resolve this
controversy.

\section{Using the Infra-Red: The PSCz survey}
\label{s:ir}
\label{s:data}
\begin{figure}[h]
\begin{center}
\psfig{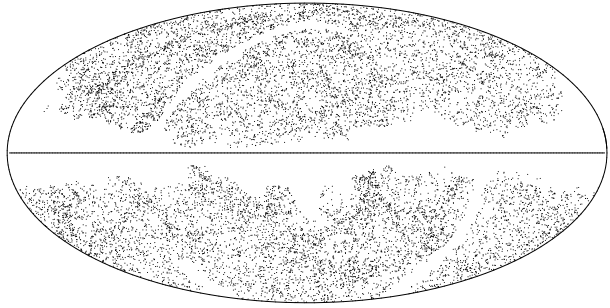}
\caption{The sky distribution of the PSCz galaxies. The line across the middle is the galactic equator.}
\label{f:psczsky}
\end{center}
\end{figure}

\begin{figure}
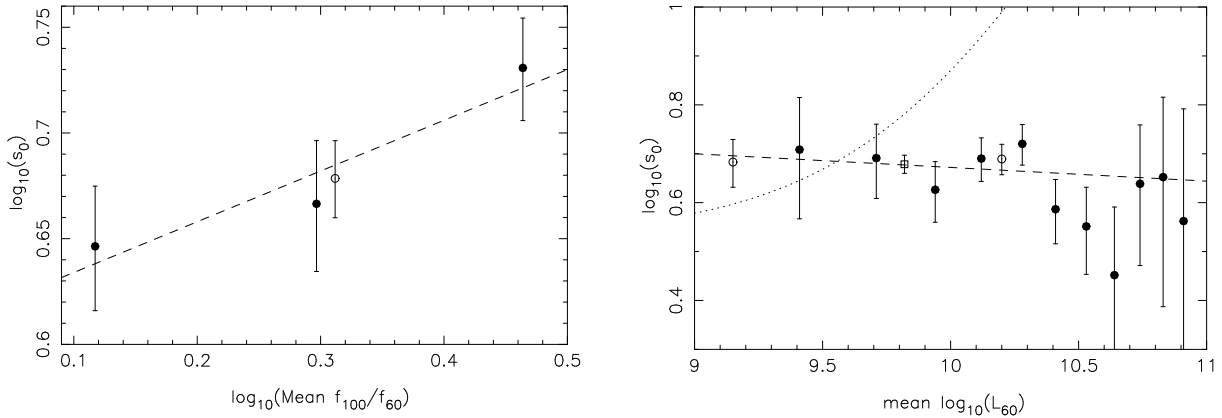

\centering
\epsfig{file=hawkins_fig2a.ps,height=.45\textwidth,clip=,angle=-90}
\hspace*{0.5cm}
\epsfig{file=hawkins_fig2b.ps,height=.45\textwidth,clip=,angle=-90}
\caption{(a) The colour sub-sample results: The filled points are for
the three colour sub-samples and the unfilled point is the result for
the whole survey. The dashed line is a best fit to the data. (b) The
luminosity sub-sample results: The filled points are the volume
limited samples, the two circles are the bright and faint sub-samples
and the square point is the result for the whole survey. The dashed
line is the fit to the colour points converted to luminosity. The dotted
line is a model prediction.}
\label{f:respscz}
\end{figure}

The publicly available PSCz survey \cite{ref3} was selected from the
\rm{\it{IRAS}} Point Source Catalogue and contains nearly 15000 galaxy
redshifts to a limiting $60\mu$m flux ($f_{60}$) of 0.6Jy. The median
redshift is $z \sim 0.03$ and the survey covers about $84\%$ of the
sky (Figure \ref{f:psczsky}). The main incompleteness is at low
galactic latitudes and is due to galactic extinction which is masked
out in the analysis. The redshift-space correlation length, $s_0$, is
measured using standard methods for the whole survey and for various
sub-samples of the catalogue.

Galaxies with a high star formation rate (SFR) tend to be bright
in the far infra-red (FIR) because of the thermal emission from dust
heated by young stars. They will hence have warmer FIR colours and so
splitting the survey by colour will test the dependence of clustering
on star formation rate.

Figure \ref{f:respscz}a shows the results for three sub-samples
selected on their $f_{100}/f_{60}$ colour. The dashed line is a best
fit to these results and it is clear that the clustering amplitude is
less for the galaxies with a lower $f_{100}/f_{60}$ ratio, and hence
warmer colour. This implies that these warmer, more star-forming
galaxies are found outside the dense regions of clusters which is
confirmed by the observations that later type optical galaxies are
found less clustered than earlier galaxy types.

The results for two luminosity selected samples and a range of
volume-limited samples are plotted in Figure \ref{f:respscz}b. The
best fit to the colour sample results is converted to a luminosity
prediction via the observed PSCz colour-luminosity relation. This is
the dashed line in Figure \ref{f:respscz}b and it can be seen that it
is consistent with the data points. The dotted line is the prediction
of a typical galaxy formation model described above and is clearly
inconsistent with the data. The conclusion is that the assumption made
in the galaxy formation models that galaxy luminosity is related to
galaxy mass is not valid in the far infra-red, and that a galaxy's far
infra-red luminosity is dominated by its SFR \cite{ref6}.

\section{Using the Optical: The 2dF Galaxy Redshift Survey}
\label{s:op}
The 2dF Galaxy Redshift Survey was selected in the $b_J$ band from the
APM survey and has a limiting apparent magnitude of $b_J = 19.45$. The
survey covers approximately 2000 square degrees and is split into two
broad strips, one in the South Galactic Pole region (SGP) and the
other in the direction of the North Galactic Pole (NGP). The aim of
the survey is to observe a total of 250,000 galaxies and as the survey
is still on-going, the results that follow are from an interim
study. To date over 170,000 unique galaxy redshifts have been
measured, with a median redshift of $z \sim 0.1$. The survey is
described in detail in \cite{ref4}, and the preliminary data release
is available at {\tt http://www.mso.anu.edu.au/2dFGRS/}. The large
number of galaxies, over a large area of the sky, makes the survey
ideal for measuring the large scale structure of the Universe. It is
also large enough to allow accurate estimates of the clustering to be
measured for sub-samples of the data, adding extra dimensions to the
output of the survey. It is this fact which makes it ideal for looking
for the dependence of clustering on galaxy properties.

The clustering is measured \cite{ref5} using the projected correlation
function, $\Xi(\sigma)$, which is obtained by projecting $\xi(\sigma,
\pi)$ onto the $\sigma$ axis, giving a correlation function
measurement free from redshift-space distortions. The real-space
correlation length, $r_0$, can then be measured from $\Xi(\sigma)$.

\begin{figure}
\plotone{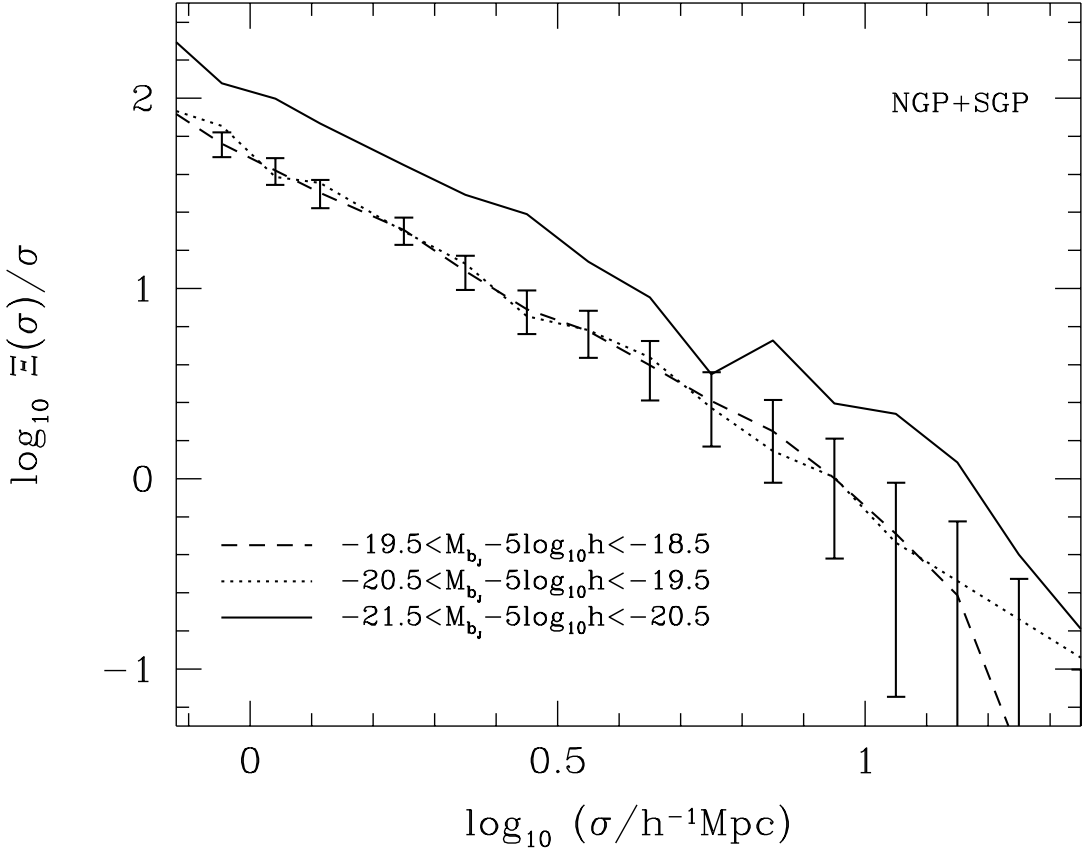}{15cm}
\caption{The projected correlation function for a single volume
limited sample split into 3 disjoint magnitude slices. The
brightest bin shows a clear increase in clustering but all have a
similar slope.}
\label{f:2df2}
\end{figure}

Figure \ref{f:2df2} shows the projected correlation function results
for a single volume limited sample defined to include all galaxies
with $-21.5 < M_{b_J} - 5\log h < -18.5$. This sample is further split
into three disjoint one magnitude slices. The brightest bin (solid
line) clearly shows stronger clustering than the fainter two bins over
a wide range of scales. The error bars come from the scatter between
mock catalogues. Another feature of this plot is that the slopes of
the measured correlation functions are very similar. Using a single
volume is an important robustness test as the volume probed is the
same for each sample and hence the cosmic variance is the same for
each line. The disadvantage is that the brighter bins contain a lot
fewer galaxies.

To overcome the lack of galaxies in the brighter bins of a single
volume, a series of half-magnitude wide volume limited samples are
used. They are created by changing the absolute magnitude limits to
maximise the number of the very bright galaxies in the brightest
bins. Thus the clustering can be measured as a function of median
absolute magnitude. The disadvantage here is that different volumes
are then being analysed and so could introduce cosmic variance.

\begin{figure}
\plotone{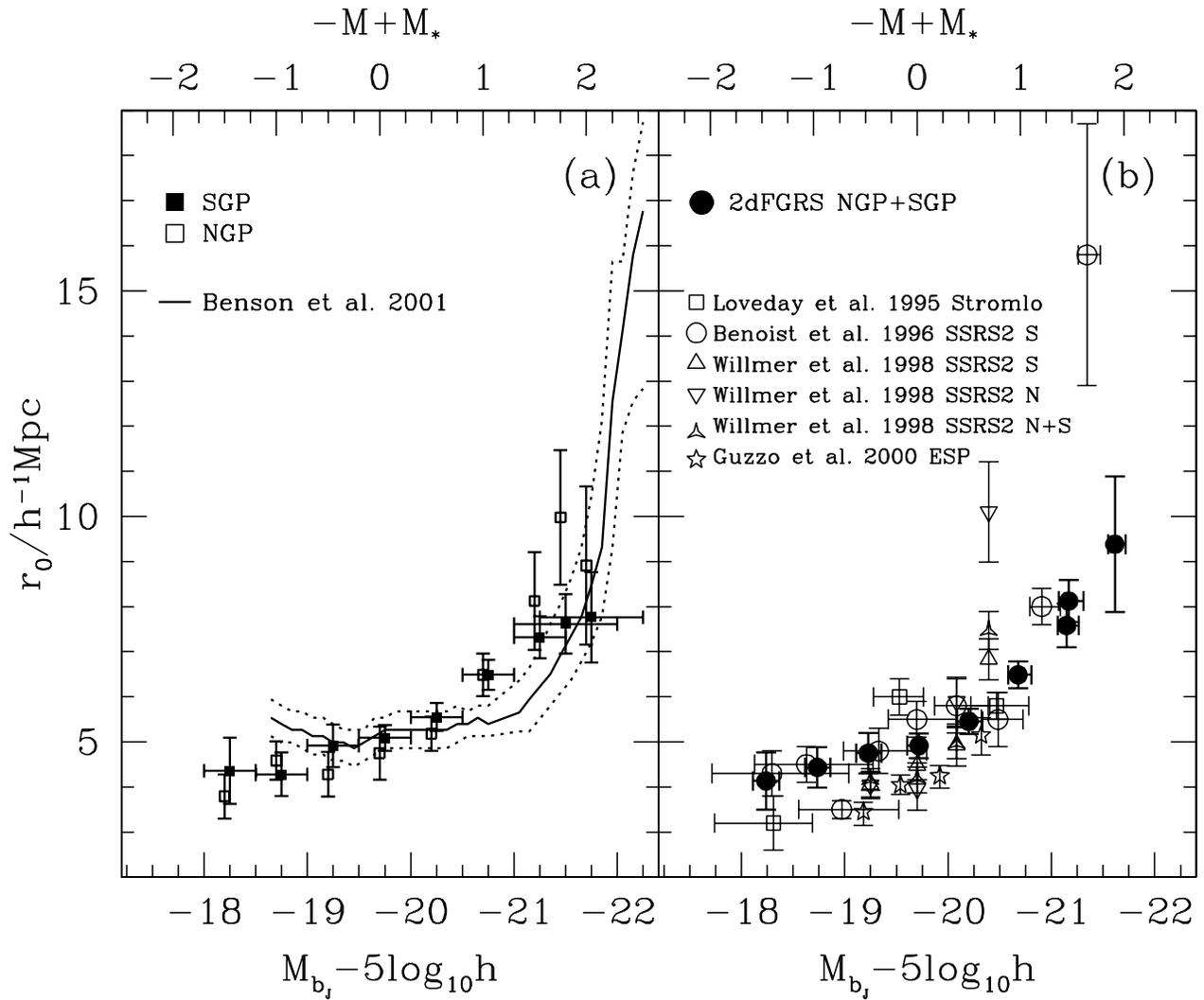}{17cm}
\caption{(a) The $r_0$ measurements for the two independent parts of
the survey compared with a semi-analytical model. They are consistent
within the errors. (b) The 2dFGRS results compared with the results
from other surveys. The error bars are from the scatter between mock
catalogues.}
\label{f:2df1}
\end{figure}

Figure \ref{f:2df1}a shows the estimates of $r_0$ measured
independently in the NGP and the SGP for the range of volume limited
samples. The results are consistent within the errors and is another
important test of the robustness of the clustering effect. In Figure
\ref{f:2df1}b the NGP and SGP measurements have been combined to give
the overall 2dFGRS results and these are compared to the results from
other surveys. The results confirm the rapid increase of clustering
for samples with $M$ brighter than $M^*$. The error bars are smaller
than previous measurements even though the scatter between mock
catalogues is used. The Poisson and bootstrap re-sampling methods of
previous surveys will tend to underestimate the errors.

\begin{figure}
\plotone{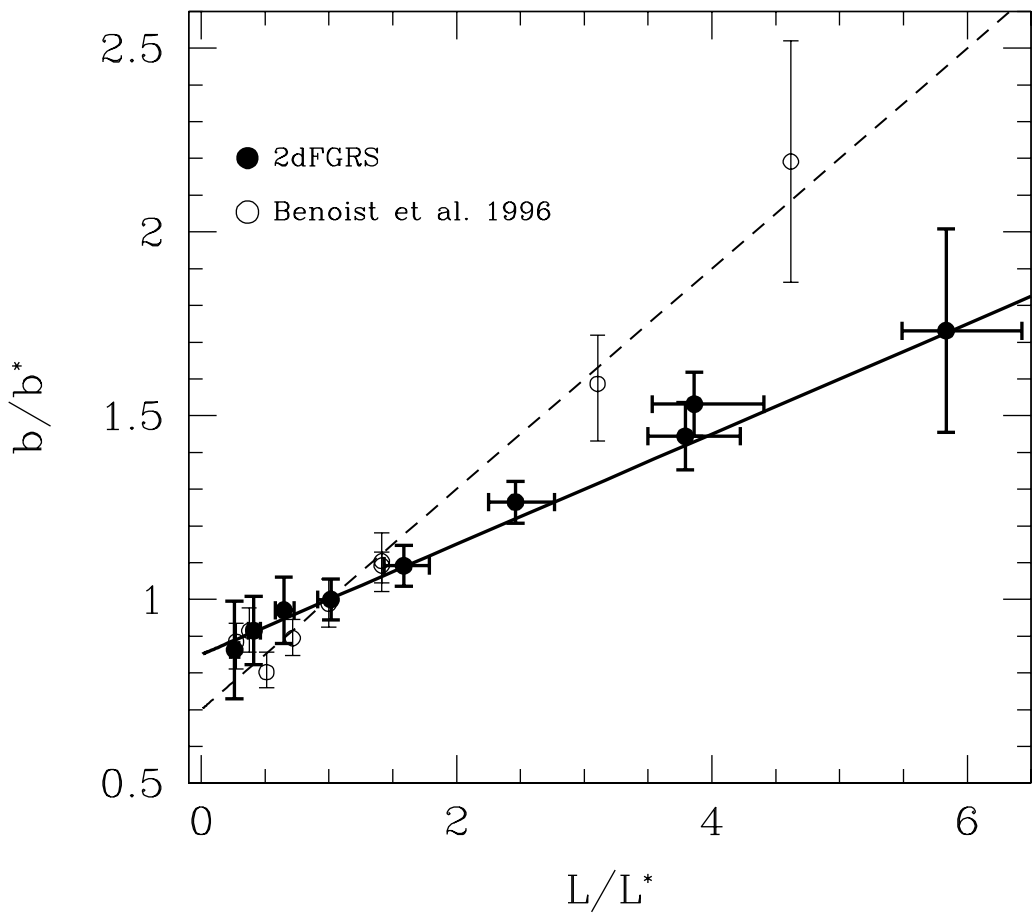}{14cm}
\caption{The relative bias as a function of relative luminosity for
the 2dFGRS (solid points) compared with the results of \cite{benoist}
(unfilled points), which have been fit by the dashed line. The 2dFGRS
results have a shallower slope and are fit by the solid line. The
error bars are from mock catalogues.}
\label{f:2df3}
\end{figure}

Although the increase in clustering appears sharp in Figure
\ref{f:2df1}, Figure \ref{f:2df3} shows the relative bias, $b/b^* =
(r_0/r_0^*)^{0.5}$, which is measured as a function of luminosity and
compared to the results of \cite{benoist}. The 2dFGRS results are
actually well fit by the linear relation, \cite{ref5}
\begin{equation}
b/b^* = 0.85 + 0.15L/L^*.
\end{equation}

\section{Summary}
\label{s:summ}
$\bullet$ Cold \rm{\it{IRAS}} galaxies are found to be more clustered
than hot \rm{\it{IRAS}} galaxies.\\$\bullet$ Bright and faint
\rm{\it{IRAS}} galaxies are found similarly clustered, consistent with
the colour relation.\\$\bullet$ Bright optical galaxies are found to
be significantly more clustered than faint optical galaxies and the
measurement presented is the most accurate yet made.

\acknowledgements{I thank the rest of the PSCz and 2dFGRS teams for all their dedicated work.}

\vfill
\end{document}